\documentclass[referee]{aa} 
\usepackage{epsfig,deluxe}
\begin{document}

\newcommand{\gsim}{\hbox{\rlap{$^>$}$_\sim$}}
  \thesaurus{06;  19.63.1}    
%
    \title{Soft Gamma-ray Repeaters and Anomalous X-ray Pulsars: 
           Magnetars or  Quark Stars?} 
    \author{Arnon Dar$^1$ and A. De R\'ujula$^2$}  
    \institute{1. Department of Physics, Technion, Haifa 32000, Israel\\ 
              2. Theory Division, CERN, CH-1211 Geneva 23, Switzerland\\} 
    \titlerunning{Soft Gamma-ray Repeaters ...} 
    \maketitle

\begin{abstract}  

Recent measurements of the spin-down rates of soft gamma ray repeaters
(SGRs) and anomalous X-ray pulsars (AXPs) have been interpreted as
evidence that these objects are ``magnetars'': neutron stars spinning down
by magnetic dipole radiation, but with a magnetic field two orders of
magnitude larger than that of ordinary neutron stars. We discuss the
evidence disfavouring this interpretation. We argue that, instead, the
observations support the hypothesis that SGRs and AXPs are neutron stars
that have suffered a transition into a denser form of nuclear matter to
become, presumably, strange stars or quark stars.
\end{abstract}

\section{Introduction} 
Consider a neutron star (NS) of radius R, with a magnetic field
${\rm B_p}$  at the magnetic poles, spinning with a period P.
 If the star's magnetic
moment is misaligned with the spin axis by an angle $\alpha$,
electromagnetic energy is emitted at a rate 
(see, e.g., Shapiro \& Teukolsky 1983 and references therein.)  
\begin {equation}  
{\rm \dot E = -\,{B_p^2\, R^6\,\Omega^4 \, \sin^2\alpha\over 6\,c^3}}\; ,
\end{equation}  
with $\Omega=2\pi/{\rm P}$ the angular velocity.
If the NS is observed as a pulsar, its period can be measured as
a function of time. Let the NS have a moment of inertia I and 
let ${\rm \dot P}$ be the rate
at which its period decreases. The magnetic field required to explain
the slow-down rate by magnetic dipole radiation (MDR) is: 
\begin{equation}  
{\rm 
B_p=\left({6\,c^3I\over 4\,\pi^2\,R^6\,  
\sin^2\alpha}\right)^{1/2}~\sqrt{\dot P\,P}}\, .
\label{field}
\end{equation} 
For a ``canonical'' NS, 
${\rm M\sim 1.4~M_\odot}$, R $\sim$ 10 km   and 
${\rm I\sim 10^{45}~g~cm^2}$, so that   
${\rm B_p=6.4\times 10^{19}}$ ${\rm \sin^{-1}\alpha\,\sqrt{\dot P P_s}}$ 
Gauss, with ${\rm P_s}$ the period in seconds.
Young radio pulsars such as the Crab pulsar are fast-rotating NSs 
with an estimated surface magnetic field ${\rm B_p\!\sim \!10^{13}}$  
Gauss. They fit the picture whereby their observed slow-down rates
are due to the MDR of their rotational energy${^{^1}}$.  

All observed slowly-rotating pulsars, of period ${\rm P\!>\!5}$ s, 
exhibit very peculiar 
properties. Four of them are known to emit occasional short bursts 
of radiation peaking at tens of keV energies,
 and are classified as ``Soft Gamma-ray Repeaters'' (SGRs).
Six other ``Anomalous X-ray Pulsars'' (AXPs)
persistently emit X-rays, but they are quiet at radio wavelengths.
Some of these objects'
observational parameters
are listed in Tables I and II.  Their rotational
energy loss  ${\rm \dot E_{rot}=I~\Omega~\dot\Omega}$ is insufficient  
to power
their observed radiation, suggesting the magnetic field energy 
${\rm E_m\sim B_p^2~R^3/12}$ 
as an alternative source. 
Both SGRs and AXPs are located in directions close to those of
supernova remnants (e.g., 
Cline, et al. 1982;  Kulkarni \& Frail 1993; Vasisht et al. 1994;  
Hurley, K. et al. 1999;
Vasisht et al. 1997a; Vasisht et al. 1997b;
Gaensler, Gotthelf \& Vasisht 1999) 
that are observable for
only tens of thousands of years. The association with SNRs
would imply that these peculiar stars are
too young to have spun-down by MDR to their
current periods, if they were born rotating as fast as
the Crab pulsar, the Vela pulsar or any other young
pulsar (the ones also associated with SNRs),  and if they also have
characteristic radio-pulsar magnetic fields:  
${\rm B\!\sim\! 10^{13}}$ Gauss.  

The values of P and ${\rm \dot P}$ for SGR 1806$-$20 and SGR 1900+14 were
recently measured (Kouveliotou et al. 1998; Kouveliotou et al. 1999; Woods
et al. 1999) and are listed in Table I. They imply, by use of
Eq. 2, magnetic fields in excess of $10^{15}$ Gauss. With the
information displayed in Table II, similarly large fields can be deduced
for AXPs (e.g., Gotthelf et al. 1999; Israel et al. 1999). Magnetic fields
of this enormous intensity can explain the rapid slow-down of SGRs and
AXPs and can store enough energy to power their emissions during their
active lifetime (Duncan and Thompson 1992; Thompson and Duncan 1995;
Thompson and Duncan 1996). Not surprisingly, the discovery of SGRs and
AXPs with fast spin-down rates was reported as the observational discovery
(Kouveliotou et al. 1998; Kouveliotou et al. 1999) of hypermagnetized
neutron stars, or {\it magnetars} (Duncan and Thompson 1992; Thompson and
Duncan 1995; Thompson and Duncan 1996).

In this paper we discuss the observational evidence implying that SGRs and
AXPs are not magnetars (Dar 1999a; Marsden et al. 1999; Dar 1999b).  We
contend that the mechanism producing their observed rapid slow-down is not
MDR, but relativistic particle emission along the magnetic axis, be it in
the form of jets (Dar 1999b) or of winds (Harding et al. 1999). We argue
that the locations and estimated ages of SGRs, when compared to those of
their associated SNRs, strongly suggest that ``something'' happened to
these NSs well after they were born. Their inferred X-ray emitting surface
areas, significantly smaller than those of a ``canonical'' neutron star,
point in the same direction.  For the source of the emitted energy we do
not have an explicit model; we conjecture that the energy gained by steady
gravitational contraction can power both the quiescent X-ray emission and
the star quakes that produce ``soft'' gamma ray bursts (Ramaty et al.
1980). A phase transition from a conventional neutron star into a strange
star or a quark star (Dar 1999a) can explain, we shall argue, all of the
properties of SGRs and AXPs.

\section{Critique of the magnetar model} 
 
The magnetar model of SGRs cannot explain their ages, locations
and occasional increases in spin-down rate (Dar 1999a, Marsden et al. 1999).
The ages of SGRs, if estimated from their magnetic spin-down rate, are 
much smaller than the ages of the remnants of the supernovae in 
which they were born: an {\it age crisis}. The location of SGRs relative to
the centre of their associated remnant implies that they move with  
unacceptably
large peculiar velocities: a {\it separation crisis.} Sudden increases  
in ${\rm \dot P}$
require inexplicable jumps in the energy stored by the magnetic field: an
{\it energy crisis.}

A pulsar with initial spin period  ${\rm P_i}$
and constant moment of inertia, whose
rotational kinetic energy ${\rm E_{rot} =I~\Omega^2/2}$
powers the MDR, has an age, t, shorter than the ``characteristic  
age'', ${\rm \tau_s}$:
\begin{equation}  
{\rm  t = {P\over 2\,\dot P}\left(1-{P_i^2\over P^2}\right) 
\leq \tau_s \equiv {P\over 2\,\dot P}}\; . 
\label{age}
\end{equation}  
An age estimate independent from the above ``magnetic braking'' age 
is provided by the time elapsed since
the parent supernova event took place. The ejecta from 
SNe expand freely to a radius ${\rm R_{SNR}\propto t}$, as long as the 
mass of the swept-up ambient medium 
is smaller than the mass of the ejecta. When they are comparable, the
SNR enters a ``Sedov-Taylor phase'' during which ${\rm R_{SNR}\propto 
t^{2/5}}$. Finally, for a swept up mass superior to the ejected mass,  
the SNR 
cools radiatively and ${\rm R_{SNR}\propto t^{2/7}}$.  The expansion 
velocity and the size of SNRs, as well as their X-ray temperatures, are 
commonly used to estimate their ages (see, e.g., Shapiro \& Teukolsky 
1983 and references therein.).  

In Table I we list the characteristic  ages $\tau_s$
of SGRs, and the ages of the SNRs where they were presumably born.
For the two SGRs whose slow-down rate ${\rm \dot P}$ has been measured  
(1806-20 and 1900+14), Eq.~\ref{age} results in $\tau_s\approx  
1400,~1300$
y, respectively. This upper limit is significantly smaller than the  
estimated
age of their SNRs, which is larger than $5\times 10^3$ and $10^4$ y,
respectively. This age crisis of the magnetar model would recur
for SGR 0529-66 and 1627-41, if their slow-down rate is similar to
that of the other two known SGRs.
 
The characteristic ages of SGR 1806-20 and 1900+14 also imply a separation
crisis.  SGR 1806+20, if it was born at the centre of SNR G10.0-0.3
(Hurley et al. 1999b) and is less than ${\rm \tau_s\approx 1400~y}$ old,
must have travelled with a sky projected velocity larger than ${\rm
v_\perp\approx 5500~(D/14.5~kpc)~km~s^{-1}}$ to its present location
(Kulkarni et al. 1994). With its similar ${\rm \tau_s}$, SGR 1900+14, if
associated to SNR G42.8+0.6 (Vasisht et al. 1994), must have travelled at
${\rm v_\perp\sim 27000~(D/7~kpc)~km~s^{-1}}$, or faster, to where it is
(Hurley et al. 1999c; Hurley et al. 1999d). The same argument, applied to
SGR 0526-66 and 1627-41 at their current locations (Cline et al. 1982;
Hurley et al. 1999a), would result in lower limits of ${\rm v_\perp\sim
22000}$ and ${ \sim 4000}$ km s$^{-1}$, respectively, if their spin-down
rates and implied ages turned out to be akin to the measured ones. These
magnetar-model velocities are too large, compared to the mean observed
${\rm v_\perp\sim 350\pm 70~km~s^{-1}}$ of pulsars (e.g., Lyne and Lorimer
1994), and in particular of young pulsars such as the Crab pulsar 
(${\rm v_\perp\approx 170~km~s^{-1}}$, Caraveo and Mignani 1999) and the
Vela pulsar (${\rm v_\perp\approx 70~km~s^{-1}}$, Nasuti et al. 1997).

In the magnetar model of SGRs the radiation-energy source 
is the magnetic field energy 
${\rm E_M \sim B_p^2\,R^3/12}$ $\sim 10^{47}$ erg, for 
${\rm B_p\!=\!10^{15}}$ Gauss and  ${\rm R\!=\!10}$ km.
Magnetic braking implies that the pulsar's surface field is  
${\rm B_p^2\propto \dot P}$, an increase in ${\rm \dot P}$  implies a  
commensurate
increase in magnetic energy. The 
spin-down rate of SGR 1900+14 roughly doubled from  
${\rm \dot P\sim 6\times 10^{-11}}$ to ${\rm \dot P\sim 13\times  
10^{-11}}$  
around the time of its large flare on 27 August 1998
(Woods et al. 1999a, Marsden et al. 1999). 
How to explain a sudden doubling of a huge magnetic energy?
This is the energy crisis. As the magnetic energy is consumed
and the field weakens, the pulsar's spin-down rate should decrease,
countrary to observation: yet another problem for the magnetar
scenario (Marsden et al. 1999). 

For AXPs the magnetar model faces similar difficulties. AXPs 1709-40 and
1E 1048-5937 have spin-down ages (9 ky and 4.6 ky) shorter than the
estimated age of their associated SNRs (20 ky and 10 ky), hinting at an
age crisis.  The projected sky velocities required to move these objects
from the centres of their associated SNRs (G346.6-0.2 and G287.8-0.5) to
their observed positions are 2100 km s$^{-1}$ and 2300 km s$^{-1}$, a
separation crisis. Observed jumps in the spin-down rate of AXPs
1E1048.1-5937 and 1E2259+58, akin to the one in SGR 1900+14, entail an
energy crisis.  For AXP 1E 2259+586 (Corbet et al. 1995), the
magnetic energy inferred from its spin-down rate, ${\rm E_B\approx 2\times
10^{45}}$ erg, is insufficient to power its steady X-ray luminosity, ${\rm
L_X\approx 8\times 10^{34}~erg ~s^{-1}}$, over its characteristic age,
${\rm \tau_s\sim 1.5\times 10^5~y}$. Also, a magnetic field this large
would be inconsistent with the absorption features observed by ASCA in its
X-ray spectrum (Corbet et al. 1995), if interpreted as cyclotron lines. 
 
The magnetar model of SGRs and AXPs is not successful: alternatives are
called for.

\section{Spin-Down by Relativistic Jets}  
 
There is evidence for the emission of relativistic particles  
by SGRs. 
In the case of SGR 1806-20, the
non-thermal quiescent X-ray emission and the highly suggestive radio 
images (Vasisht et al. 1995; Frail et al. 1997) provide 
compelling evidence for steady relativistic particle emission,
perhaps in the form of relativistic jets.  A 
fading radio source is seen within the 
localization window of SGR 1900+14; it has 
been interpreted as a short-lived nebula 
powered by relativistic particles ejected during the intense high energy 
activity in late August 1998 (Frail et al. 1999). The 
emission of relativistic particles along the magnetic axis can be the 
dominant mechanism for the braking of slowly-rotating 
pulsars with normal magnetic fields (Dar 1999b, see also Harding et al.
1999), for which magnetic braking is  
inefficient. Magneto-hydrodynamic calculations 
of pulsar braking by particle emission are a formidable task, but simple   
estimates (Dar 1999b) will suffice here.  
 
Let L$_{\rm RP}$ be a pulsar's luminosity in the form of relativistic
particles escaping from the magnetic poles along the open 
magnetic lines. The emitted particles co-rotate with the magnetic field
up to a radius ${\rm r_e \approx (3\,c\,B_p^2\,R^6/2\,L_{RP})^{1/4}}$,  
at which
their pressure (${\rm L_{RP}/[12\pi\,c\,r^2}]$) becomes comparable to  
the magnetic
pressure (${\rm B_p^2\,R^6/[8\pi\,r^6]}$ for a dipole field). 
Beyond this point a particle
of mass m and Lorentz factor $\gamma$ is no longer entangled
in the magnetic field and it escapes, carrying away an angular momentum
${\rm \gamma\,m\,\Omega\,r_e^2\,\sin^2\alpha}$. The resulting rate of
rotational energy loss is 
\begin{equation}
{\rm \dot E_{rot}=I\,\Omega\,\dot\Omega 
\approx -  
\left( {3\,L_{RP}\over 2\,c^3}\right)^{1/2}\,B_p\,R^3\,\Omega^2 
\,\sin^2\alpha},
\end{equation} 
which yields an exponential 
decline, ${\rm E_{rot}(t)=E_{rot}(0)\,exp(-t/\tau_s)}$, with a 
characteristic time: 
\begin{equation}  
{\rm \tau_s={P\over 2\,\dot P}={I\over B_p\,R^3\,\  
sin^2\alpha}\left({c^3\over  
6\,L_{RP}}\right)^{1/2}}.  
\end{equation} 
For a conventional ${\rm B_p=10^{13}}$ Gauss, ${\rm R=10}$ km,
${\rm L_{RP}=10^{37}}$ erg s$^{-1}$, 
and $\sin\alpha\approx 1$, the characteristic slow-down time is  
${\rm \tau_s\sim  2000~y}$, scaling as 1/R at fixed  ${\rm B_p\,R^2}$,
and consistent with the characteristic slow-down times of SGRs and AXPs.  

Relativistic particle emission may also be the dominant spin-down mechanism
in pulsars rotating faster than SGRs and AXPs.
A comparison between Eq.4 and Eq.1 shows that slowing-down by relativistic 
jets becomes faster than slowing-down by MDR when 
\begin{equation}
{\rm P\geq {(4\pi^2\,B_p\,R^3)^{1/2}\over (54\,c^3\,L_{RP})^{1/4}}}.
\end{equation}
For ${\rm B_p=10^{13}}$ Gauss, ${\rm R=10}$ km,
and ${\rm L_{RP}=10^{37}}$ erg s$^{-1}$, 
slowing-down by emission of relativistic jets is faster than by MDR if 
${\rm P> 60~ms}$. But, for
${\rm B_p=10^{13}~ Gauss}$, ${\rm L_{RP}<10^{37}~erg}$, 
and ${\rm P<100~ms}$, 
${\rm r_e\,\sin\alpha}$ 
becomes larger than the radius of the light cylinder, ${\rm r_c=c/\Omega}$,
and  
relativistic particles of energy E that stop co-rotating with the pulsar 
at ${\rm r_c}$ carry away an angular momentum ${\rm E/\Omega}$,
so that the total rate of angular momentum loss by particle emission is
${\rm \dot L\approx L_{RP}/\Omega} $, i.e., 
\begin{eqnarray}
{\rm \dot  E_{rot}\approx L_{RP}, \qquad}
{\rm \tau_s\approx E_{rot}/L_{RP}}.
\end{eqnarray}
The relation ${\rm \dot E_{rot}=L_{RP}}$ is well  
satisfied, for instance, by the Crab pulsar,  for which 
${\rm \dot E_{rot}=I\,\Omega\,\dot\Omega\approx}$ ${\rm 5\times 
10^{38}~erg~s^{-1}}$, coinciding exactly with the estimated energy
input to the Crab nebula (Manchester and Taylor 1997), presumably 
supplied by relativistic particles from the pulsar.  
 
The gamma-ray bursts and radio flares of SGRs are presumably produced by
bursts of relativistic particles.
If relativistic particle emission induces
the observed spin-down, ${\rm \dot P}$ should increase during these periods
of activity.  Indeed, the
spin-down rate of SGR 1900+14 doubled during its intensive burst 
activity in 1998, after which it seems to resume
 its ``quiescent'' long-term value (Woods et al. 1999a), as shown in  
Fig.~1.

\section{What powers SGRs and AXPs?} 

If, unlike in the magnetar model, the energy reservoir  of SGRs and AXP
is not magnetic, what can it be? 
A NS whose internal heat, magnetic field and/or angular momentum 
are diminishing as it radiates, may undergo a phase transition
(see, e.g., Shapiro \& Teukolsky 1983 and references therein) 
and collapse to a strange star (SS) or a quark star (QS).
Gravitational energy release during the subsequent
slow contraction of the cooling and spinning-down star may 
power  SGRs and AXPs (Dar 1999a).  
The equation of state of nuclear matter, or even that of quark matter at
supernuclear densities, has not yet been derived from first principles.
Yet, simple considerations indicate
that the possible phase transitions of NSs into SSs and QSs ought to  
be taken
seriously.

Naively approximate the pressure of cold nuclear matter at NS   
densities by that
of a non-relativistic degenerate Fermi gas of nucleons. Ignoring
general-relativistic corrections, the radius and central  
density ${\rho_c}$ of a self-gravitating gas of neutrons of  
total baryonic mass M  and zero angular momentum  
are then given by the polytropic Emden-Lane solution of the  
hydrostatic equation:
\begin{equation} 
{\rm R\approx 15.1 
       \left({M\over M_\odot}\right)^{-1/3}~km},  
\end{equation} 
 \begin{equation} 
{\rm \rho_c\approx 6\,\bar\rho \approx 0.83\times 10^{15} 
             \left({M\over M_\odot}\right)^2~g~cm^{-3}}.  
\end{equation} 
In this simplest of models, low mass NSs should indeed be made of neutrons,
but as M is increased past ${\rm 1.27\,M_\odot}$, 
${\rm \rho_c}$ increases until the central Fermi energy
${\rm E_F=(h^2/8\,m_n)(3\,\rho_c/\pi\, m_n)^{2/3}}$ exceeds 
${\rm (m_\Lambda-m_n)\,c^2}$. At this point, it is favourable for the
strangeness changing weak process  ${\rm n\to\Lambda}$ (or ${\rm 
ud\rightarrow su}$) to start transforming neutrons at the top of the  
Fermi sea 
into (initially pressureless) $\Lambda$'s at the bottom of the sea.
This reduces the pressure, causes contraction and increases ${\rho_c}$,
initiating a run-away reaction that stops only as the n and $\Lambda$
chemical potentials equalize, i.e. 
until ${\rm E_F(n)-E_F(\Lambda)\approx}$ 
${\rm  c^2\,(m_{_\Lambda}-m_n)(1-GM(r)/r)}$, 
where ${\rm M(r)}$ is the mass enclosed within r.

At the central densities of the strange stars of the previous paragraph,
the nucleons would be so snuggly packed that their ``individuality'' would
be in doubt.
But it has been argued (Alford et al. 1998; Berges et al. 1999
Rapp et al. 1999; Wilczek 1999; Li et al. 1999)  that cold 
nuclear matter,  
compressed to high nuclear densities, converts into a much denser
superfluid and superconducting Bose-condensate of spin zero diquarks.
Cooper pairing of quarks 
reduces their pressure and would trigger a gravitational collapse that,
if it does not proceed all the way to a black hole, would stop
only when the squeezed size of the pairs increases their internal 
energy above their binding energy.  

A neutron star may be born with a temperature, a magnetic field and/or
an angular momentum that prevent its transition to a strange- or  
quark-matter
state. As the star ages, it may reach a point at which a transition to
a denser state of matter is favourable. The collapse would
reheat the star to some extent; we conjecture that the gravitational energy
made available by its subsequent slow cooling and contraction 
can power SGRs and AXPs (estimates of the effect are difficult, since  
quantities
such as the heat conductivity are notoriously hard to predict).
For a pulsar which is mainly supported by the Fermi pressure of 
non-relativistic degenerate fermions contraction can power a total
luminosity:
\begin{equation}  
{\rm 
L\approx {2\over 7}\left({G\,M^2\over R}\right) {\dot R\over R}
}\; . 
\end{equation}  
For a canonical pulsar mass ${\rm M=
1.4~M_\odot}$ and a radius ${\rm R=10}$ km, a contraction rate of  
${\rm \dot 
R \sim 20~\mu m~ y^{-1}}$ (a tiny ${\rm \dot R/R\sim 2\times  
10^{-9}~y^{-1}}$)
is sufficient to provide the inferred total luminosity of 
SGRs and AXPs, ${\rm L \leq 10^{37}~erg~s^{-1}}$ .  

\section{Extra evidence and hints in favour of collapsed NSs} 
 
The gravitational collapse of 
pulsars to strange or quark stars may offer explanations for some 
puzzling observations: the anomalously small effective surfaces of AXPs,
the origin of short duration gamma-ray bursts, the shape of some SNRs 
and the large peculiar velocities of old pulsars.

The X-ray spectra of SGRs and AXPs in quiescent periods have 
been interpreted as the Wein tail of black-body 
radiation from their surface. The Stefan-Boltzman law, ${\rm L_X=4\pi\, 
R^2\sigma\, T^4}$ (or ${\rm L_X \approx 1.3\times 10^{37}~erg~s^{-1}}$, 
for ${\rm R=10}$ km and ${\rm T=1}$ keV)
yields effective surface areas significantly 
smaller than expected for a  NS, ${\rm A_{NS}\approx 4\pi\times 10^2~km^2}$.
The AXP data are summarized in Table II, 
normalized to the measured distances  (analysis of the corresponding 
data for SGRs is complicated by their time variability and not well
determined temperatures).
All inferred areas are $\sim$20\% 
of the expectation. It would be difficult to attribute this
systematic effect to errors in the observations.
Effective areas smaller than expected 
may be due to non-uniform surface temperatures. But, more interestingly,
they can be real and reflect the small radii of SSs or QSs.  

Does the transition from a neutron star to a denser star
have directly observable signatures?
The answer 
may be guided by analogies with observed phenomena,
a detailed model would be very hard to develop. 
The gravitational binding energy release --of ${\cal O}(10^{53})$ ergs--
would be mainly emitted as a neutrino burst,
as in the Type II explosion that first begat the NS.
The collapse of a NS core
into a denser object should be accompanied by the ejection
of the outer layers, and be more similar to a Type I SN explosion
than to Type II, Ib or Ic events,  for which the ejected mass is much 
larger and consists mainly of light elements.
The ejecta should be mildly relativistic and deposit their
energy in the interstellar medium at a fast rate, giving rise to a  
short-lived
SNR, rich in Fe-group elements. 
The collapsing material may, as in active galactic nuclei,
 acquire an accreting toroidal structure and emit
highly relativistic and collimated jets. These jets, if they 
point in our direction, may produce 
gamma-ray bursts (Dar 1999a; Dar and Plaga 1999)

If collimated jets produce cosmological gamma-ray bursts,  
their kinetic energy must be ${\rm E_k\sim 10^{52}~erg}$, i.e. comparable 
to the kinetic energy of the SNR from the SN event in which the NS
was originally born. With that much energy, the jets may 
distort the first SNR in a  recognizable manner.  
Radio observations expose a vast range of SNR shapes
(see, e.g., Whiteoak and Green 1996). While very 
young SNRs have a simple expanding geometry, most older SNRs 
have a distorted and 
complicated appearance, which has been traditionally attributed 
to their expansion into an inhomogeneous interstellar medium. 
However, some SNRs have striking properties which require  
explanation either in terms of jets
(Manchester 1987; Rozyczka 1993; Gaensler 1998) or --if not due to
accidental superpositions-- in terms of  a second
explosion (Aschenbach  1998; Aschenbach et al. 1999; Gaensler 1999).
A second gravitational collapse, which produces 
a second bang and emits relativistic jets along the rotation axis may 
explain the puzzling morphology of many SNRs (Dar and De R\'ujula, in
preparation). 

In the collapse of a neutron star, an imbalance in the momenta of
oppositely ejected jets can impart a natal kick to the resulting SS or QS. 
This may explain the large observed velocities of ``old'', slowly-rotating
pulsars.  Millisecond pulsars and young pulsars have small velocities
(e.g., Toscano et al. 1999); their youth and large angular momentum may
have temporarily prevented their collapse (for millisecond pulsars this
may also be a selection effect: they are found in binary systems and only
with a small natal velocity could they remain bound and be spun up by mass
accretion).


\section{Outlook} 

We have contended that SGRs and AXPs do not have the anomalously large
magnetic fields postulated in the magnetar model to be the cause of
their fast spin-down and the energy reservoir 
of their emitted radiation. 
Instead, we argued that the spin-down
is caused by relativistic particle emission and we conjectured that
the power supply is the  gravitational energy released by contraction,
resolving the conundrum associated with the large observed
jumps in spin-down rate.

Independently of the strength of their magnetic field, the well
measured SGRs and AXPs are truly puzzling: their spin-down ages
are much smaller than the age of the SNRs with which they are
associated, and the distance they must have travelled during their
lifetime implies an unacceptable velocity. The hypothesis that these
neutron stars have suffered a delayed transition to a denser type of  
constituency
resolves these problems: the measured spin-down ages  date back only to the
stars' ``second birth''. This hypothesis also explains why the star's  
surfaces,
as extracted from their X-ray emission, turn out to be smaller than expected
for a conventional NS, and why the morphology
of some of their associated 
SNRs hints at a double bang. 
If the second birth gives a new kick velocity to the NS, its direction  
should be
uncorrelated to the centre of the SNR, as in the Vela pulsar.

Most observed pulsars are not in binary systems and are not SGRs or AXPs.
These conventional pulsars have periods averaging to 1/2 s, significantly
shorter than the periods listed in Tables I and II. Their characteristic
spin-down ages, on the other hand, are longer, typically $10^7$ years. 
With such long
lifetimes, and if a good fraction of the rate of core-collapse supernova
(roughly one per century in our galaxy) results in pulsars,
one would expect to detect some $10^5$ of these objects, while only
about $10^3$ are actually observed. But, if within some $10^5$ years
a good fraction of these NSs --depending on their mass, rotation period
and magnetic field-- were to suffer a transition into a denser object, 
the observed numbers of supernovae, conventional pulsars, SGRs and
AXPs would fall into a consistent picture.
A star freshly reborn after a phase transition could be an SGR, whose  
longer period
is explained by rapid spin-down. In turn, SGRs could 
convert into AXPs after a period of bursting activity. 
As the AXPs cool and spin down, they should become slowly 
rotating, radio-quiet, X-ray-dim pulsars. Many
of these dim pulsars should still be present in the neighbourhood
of their SNRs.  Very sensitive X-ray searches are required to discover  
 their 
presence there.

\begin{acknowledgements}
{\bf This research was supported in part by the Fund For 
Promotion Of Research At The Technion.} 
\end{acknowledgements}
{}

\newpage
\footnotesize
\begin{deluxetable}{lcccccccc}
\tablefontsize{\tiny}
\tablecaption{Soft Gamma Ray Repeaters [SGRs]\vfill}
\label{tbl-1}
\tablehead{
\colhead{Pulsar} & \colhead{SNR} & \colhead{P} & \colhead{${\rm \dot P}$} &
\colhead{${\rm B_p}$} &
\colhead{${\rm \tau_s}$} &
\colhead{${\rm \tau_{_{SNR}}}$} &
\colhead{${\rm S_\perp}$}&\colhead{${\rm v_\perp}$}  \nl
\colhead{} & \colhead{} & \colhead{(s)}
& \colhead{}
& \colhead{(Gauss)}
& \colhead{(ky)} 
& \colhead{(ky)} 
& \colhead{(pc)}
& \colhead{(${\rm km~s^{-1}}$)} }
\startdata
SGR 1806-20$^b$ & G10.0-0.1  & 7.47  &$8.3\times 10^{-11}$
& $1.6\times 10^{15}$  & 1.4 & $>5$ 
& $8.3\,{\rm d_{15}}$ & $5500\,{\rm d_{15}}$ \nl
SGR 1900+14$^c$  & G42.8+0.6  & 5.16  &$6.1\times 10^{-11}$
& $1.1\times 10^{15}$  & 1.3 
&$>10$ & $36\,{\rm d_{7}}$ & $ 27000\,{\rm d_{7}}$ \nl
SGR 1627-41$^d$& G337.0-0.1  & 6.41 &\dots & \dots & \dots & $>5$
&$5.4\,{\rm d_{11}}$ &$5400\,{\rm d_{11}/\tau_3}$ \nl
SGR 0525-66$^e$ & N49    & $\sim 8$ &\dots & \dots 
&\dots & $>5$ &$>22$  & $>22000/\tau_3$\nl
\enddata
\tablenotetext{a}
{${\rm B_p}$ values are for sin$\alpha=1.$ Distances 
${\rm d_x}$ are in units of x kpc. $\tau_3$ is
the age of the SGR in ky. \\
$^b$ Atteia et al. 1987; Kulkarni \& Frail 1993; Kouveliotou 
et al. Ref. 1994; Kulkarni et al. 1994; Murakami et al. 
1994; Sonobe et al. 1994; Kouveliotou et al. 1998 and references therein.\\
$^c$ Kouveliotou et al. 1994; Vasisht et al 1994; Kouveliotou et al. 1999; 
Woods et al. 1999a; Hurley et al. 1999d and 
references therein.\\ 
$^d$ Woods et al. 1999b; Hurley et al. 1999a  and 
references therein.\\ 
$^e$ Mazets et al. 1979; Cline et al. 1982; Marsden et al. 1996 and 
references therein.} \end{deluxetable}

\begin{deluxetable}{lcccccccc}
\tablefontsize{\tiny}
\tablecaption{Anomalous X-Ray Pulsars [AXPs] \vfill}
\label{tbl-2}
\tablehead{
\colhead{Pulsar}& \colhead{P} &
\colhead{${\rm \dot P}$} & 
\colhead{${\rm B_p}$} &
\colhead{kT} & 
\colhead{${\rm L_X}$}&\colhead{${\rm A_s}$}  \nl
\colhead{}& \colhead{(s)}
& \colhead{}
& \colhead{(Gauss)}
& \colhead{(keV)} 
& \colhead{(${\rm 10^{35}~erg~s^{-1}}$)}
& \colhead{(${\rm 4\pi\times 10^2 km^2}$)} }
\startdata
1E 1841$-$045$^b$ & 11.76 & $4.1\times 10^{-11}$
& $1.4\times 10^{15}$  
&0.55& $3\,{\rm d_{7}}^2$ & $0.25\,{\rm d_{7}^2}$  \nl
1E 2259$+$586$^c$ & 6.98  & $\sim 5 \times 10^{-13}$
& $1.2\times 10^{14}$  &0.41
& $0.8 \,{\rm d_{4}^2}$ & $ 0.22\,{\rm d_{4}^2}$   \nl
4U 0142$+$615$^d$ & 8.69  & $\sim 2\times 10^{-12}$ 
& $2.6\times 10^{14}$  & 0.39
& $0.7\,{\rm d_{4}^2}$ & $ 0.24\,{\rm d_{4}^2}$ \nl
1E 1048$-$5937$^e$ & 6.44  & $1.5-4\times 10^{-11}$ 
& $>6.3\times 10^{14}$  & 0.64 
& $5 \,{\rm d_{10}^2}$ & $ 0.23\,{\rm d_{10 }^2}$ & \nl
RX J170849$-$4009$^f$ & 11.00 & $2\times 10^{-11}$
& $9.5\times 10^{14}$  & 0.40
& $10 \,{\rm d_{10}^2}$ & $0.17\,{\rm d_{10}^2}$ \nl
PSR J1844-0258$^g$ & 6.97  &    \dots 
& \dots  &0.64
& $3 \,{\rm d_{15}^2}$ & $ 0.15 \,{\rm d_{15}^2}$ \nl
\enddata
\tablenotetext{a}
{${\rm B_p}$ values are for sin$\alpha=1.$ 
All X-ray luminosities  are in the $\sim 1-10$ keV energy band
as corrected for absorption by Gotthelf and Vasisht 1998.
The distances ${\rm d_x}$ are in units of x kpc. \\
$^b$ Vasisht \& Gotthelf 1997\\ 
$^c$ Iwasawa  et al. 1992; Corbet et al. 1995; Parmar et al. 1998 
and references therein.\\
$^d$ Mereghetti \& Stella 1995; White  et al. 1996; 
Israel et al 1999 and references therein.\\
$^e$ Oosterbroek et al. 1998; Mereghetti et al. 1997; 
Corbet \& Mihara 1997 and references therein.\\
$^f$ Sugizaki et al. 1997\\
$^g$ Torii et al. 1998; Gaensler et al. 1999 and references therein.\\}
\end{deluxetable}

\newpage 
\begin{figure} 
\begin{center} 
\epsfig{file=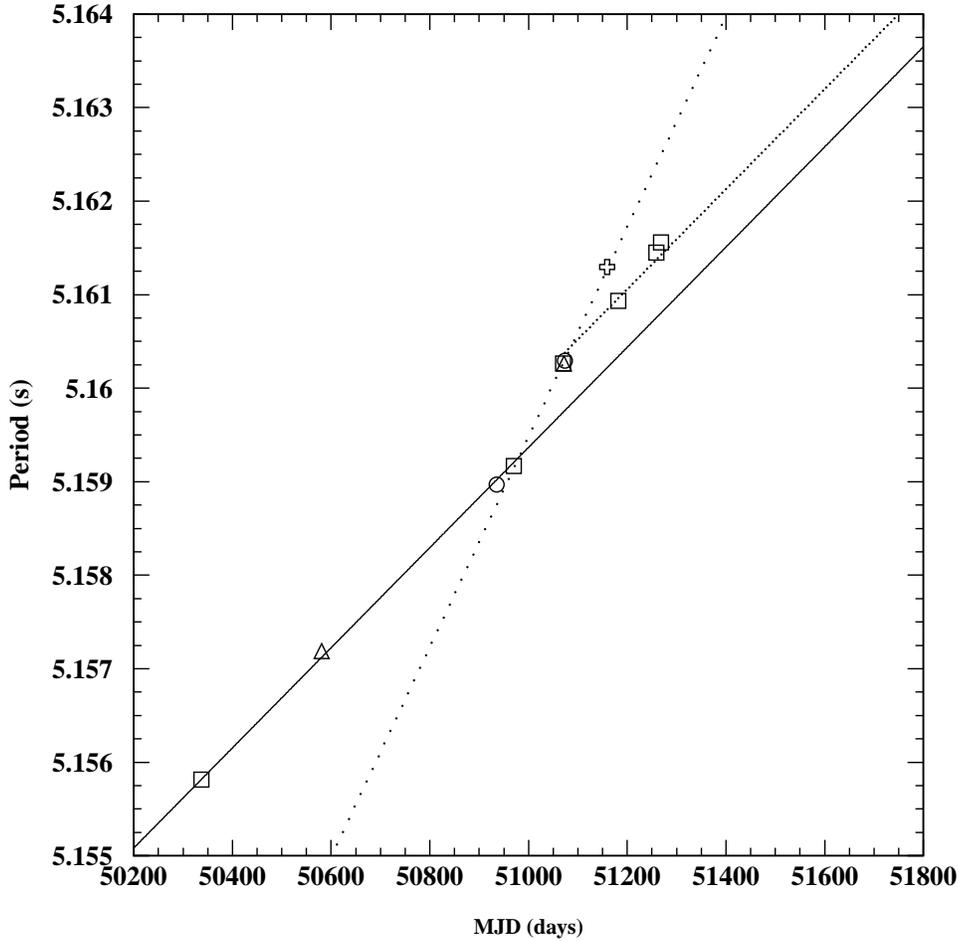,width=14cm} 
\caption{The period of SGR 
1900+14 as a function of time (Woods et al. 1999a), as measured by RXTE  
(squares), BeppoSAX (triangles), ASCA (circles) and BSA (crosses). The 
lines are linear fits to the X-ray periods of the SGR before June 
9, 1998 (${\rm \dot P=6.1\times 10^{-11}}$), between June 9 - August 27, 
1998 (${\rm \dot P=1.3\times 10^{-10}}$), and after August 27, 1998 (${\rm 
\dot P=6.1\times 10^{-11}}$).  Between June 9-August 28, the ``averaged''
spin-down rate has changed by a factor  
$\sim 2.2$ as a result of a continuous or sudden braking.}  
\vspace*{-0.5cm} 
\end{center} 
\end{figure} 

\begin{thebibliography}{}


\bibitem{Alford}
Alford, M. et al. 1998, Phys. Lett. B422, 247 
\bibitem{Asaoka} 
Asaoka, I., \& Aschenbach, B. 1994, A\&A 284, 573   
\bibitem{Aschenbach} 
Aschenbach, B. 1998, Nature 396, 141   
\bibitem{Aschenbach2} 
Aschenbach, B., Iyudin, A. F. \& Sch\"onfelder, V. 1999, astro-ph/9909415
\bibitem{Atteia} 
Atteia, J. L. et al. 1987, ApJ. 320, L105 
\bibitem{Berges}
Berges, J. \& Rajagopal. 1999, Nucl. Phys. B538, 215 
\bibitem{Caraveo}
Caraveo, P. A. \& Mignani, R. P. 1999, A\&A 344, 367 
\bibitem{Cline} 
Cline, T. L. et al. 1982, ApJ 255, L45   
\bibitem{Corbet2} 
Corbet, R. H. D. \& Mihara, T. 1997, ApJ 475, L127 
\bibitem{Corbet}
Corbet, R. H. D. et al. 1995, ApJ 443, 786 
\bibitem{Dar} 
Dar, A. 1999a, A\&A Suppl. 138, 505  
\bibitem{Dar2} 
Dar, A. 1999b, sub. to ApJ.  astro-ph/9911473 
\bibitem{Dar3} 
Dar, A. and Plaga, R. 1999, A\&A {\bf 349}, 259 
\bibitem{Duncan} 
Duncan, R. C. \&  Thompson, C. 1992, ApJ. 392, L9  
\bibitem{Frail} 
Frail, D. A., Vasisht, G. \& Kulkarni, S. R. 1997, ApJ. 480, L129
\bibitem{Frail2} 
Frail, D. A. et al. 1999, Nature 398, 127 
\bibitem{Gaensler3} 
Gaensler, B. M. 1999, {\it Ph.D. Thesis,} Univ. of Sidney (unpublished)
\bibitem{Gaensler}
Gaensler, B. M., Gotthelf, E. V. \& Vasisht, G. 1999, ApJ 526, L37 
\bibitem{Gaensler 2} 
Gaensler, B. M., Green, A. J. \& Manchester, R. N. 1998,
M. N. R. A. S. 299,  812
\bibitem{Gotthelf2} 
Gotthelf, E. V. \& Vasisht, G. 1998, New Astr.  3, 293
\bibitem{Gotthelf}
Gotthelf, E. V., Vashisht, G. \& Dotani, T. 1999, ApJ 522, L49
\bibitem{Harding} 
Harding, A. K., Contopoulos, I. \& Kazanas, D. 1999, ApJ 525, L125
\bibitem{Hurley} 
Hurley, K. et al. 1999a, ApJ 519, L143 
\bibitem{Hurley0} 
Hurley, K. et al. 1999b, ApJ 523, L37 
\bibitem{Hurley3} 
Hurley, K. et al. 1999c, ApJ 510, L111 
\bibitem{Hurley2} 
Hurley, K.  et al. 1999d, ApJ. 510, L107
\bibitem{Israel}
Israel, G. L. et al. 1999, A\&A  346, 929  
\bibitem{Iwasawa} 
Iwasawa, K., Koyama, K. \& Halpern, J. P. 1992, PASJ  44, 9 
\bibitem{Kulkarni} 
Kulkarni, S. R. \& Frail, D. A. 1993, Nature 365, 33 
\bibitem{Kulkarni2} 
Kulkarni, S. R. et al. 1994, Nature 368, 129 
\bibitem{Kouveliotou3} 
Kouveliotou, C. et al. 1994, Nature 368, 125 
\bibitem{Kouveliotou2} 
Kouveliotou, C. et al. 1998, Nature 393, 235   
\bibitem{Kouveliotou} 
Kouveliotou, C. et al. 1999, ApJ 510, L115   
\bibitem{Kouveliotou33} 
Li, X. D., {\it et al.}  1999, Phys. Rev. Lett. 83, 3776
\bibitem{Lyne} 
Lyne, A. G. \& Lorimer, D. R. 1994, Nature 369, 127
\bibitem{Manchester2} 
Manchester, R. N. 1987, A\&A 171, 205  
\bibitem{Manchester} 
Manchester, R. N. \&  Taylor J. H. 1997, {\it Pulsars,} Freeman Eds., San 
Francisco, 1997    
\bibitem{Marsdena}
Marsden, D. et al.  1996, ApJ. 470, 513 
\bibitem{Marsden}
Marsden, D., Rothschild, R. E. \& Lingenfelter, R. E. 1999, ApJ 520, L107 
\bibitem{Mazets}
Mazets, E. P. et al. 1979, Nature, 282, 587
\bibitem{Mereghetti} 
Mereghetti, S., Belloni, T. \& Nasuti, F. P. 1997, A\&A 321, 835 
\bibitem{Mereghetti2} 
Mereghetti, S. \&  Stella, L. 1995, ApJ. 442, L17 
\bibitem{Murakami} 
Murakami, T. et al. 1994 Nature, 368, 127 
\bibitem{Nasuti}
Nasuti, F. P. et al. 1997, A\&A 323, 839  
\bibitem{Oosterbroek} 
Oosterbroek, T. et al. 1998, A\&A 334, 925  
\bibitem{Parmar}
Parmar, A. et al. 1998, A\&A, 330, 175 
\bibitem{Ramaty}
Ramaty, R. et al. 1980. Nature {\bf 287}, 122 
\bibitem{Rapp}
Rapp, R. {\it et al.} 1999, Phys. Rev. Lett. 81, 53   
\bibitem{Rozyczka} 
Rozyczka, M. et al. 1993, M.N.R.A.S. 261, 674
\bibitem{Shapiro} 
Shapiro, S. L.  \& Teukolsky, A. A. 1983, Black Holes, White Dwarfs 
and Neutron Stars, {\it John Wiley \& Sons Inc. (1983)},
\bibitem{Sonobe}
Sonobe, T. et al. 1994, ApJ. 436, L23 
\bibitem{Stella} 
Stella, L., Israel, G. L. \& Mereghetti, S. 1998, Adv. Sp. Res. 22 1025
\bibitem{Sugizaki} 
Sugizaki, M. et al. 1997,  PASJ, 49, L25 
\bibitem{Thompson}
Thompson, C. \& Duncan, R. C.  1995, M.N.R.A.S. 275, 255
\bibitem{Thompson2}
Thompson, C. \& Duncan, R. C.  1996, ApJ 473, 322
\bibitem{Torii}
Torii, K. et al. 1998, ApJ. 503, 843  
\bibitem{Toscano} 
Toscano, M. et al. 1999, M. N. R. A. S. 307, 925 
\bibitem{Vasisht4} 
Vasisht, G., Frail, D. A. \& Kulkarni, S. R. 1995, ApJ. 440, L65
\bibitem{Vasisht3} 
Vasisht, G \& Gotthelf, E. V. 1997, ApJ 486, L129  
\bibitem{Vasisht} 
Vasisht, G. et al. 1994, ApJ  431, L35 
\bibitem{Vasisht2} 
Vasisht, G. et al. 1997, ApJ 476, L43  
\bibitem{White} 
White, N. E. et al. 1996, ApJ 463, L83 
\bibitem{Whiteoak}
Whiteoak, J. B. Z. \& Green, A. J. 1996, A\&A Suppl. 118, 329 
\bibitem{Woods} 
Woods, P. M. et al. 1999a, ApJ 524, L55 
\bibitem{Woods2} 
Woods, P. M. et al. 1999b, ApJ  519, L139 
\bibitem{wilczek}
Wilczek, F. 1998, Nature 395, 220  

\end{thebibliography}
\end{document}